\documentclass[sigconf]{acmart}

\usepackage{amsmath}
\usepackage{balance}

\AtBeginDocument{%
  }

\copyrightyear{2027}
\acmYear{2027}
\setcopyright{acmlicensed}
\begin{document}
% \newtheorem{theorem}{Theorem}[section]
% \newtheorem{corollary}[theorem]{Corollary}
% \newtheorem{dfn}{Definition}
% \newtheorem{hypothesis}{Hypothesis}
% \newtheorem{lemma}[theorem]{Lemma}
% \newtheorem{proposition}[theorem]{Proposition}
% \newtheorem{remark}{Remark}
% \newtheorem{example}{Example}


% \title{UniRec: End-to-End Unified Cross-stage Multi-Task Fusion for Preference Optimization in Recommender Systems}
\title{UniRec: Cross-Stage Multi-Task Fusion with Preference Optimization for Cascaded Recommender Systems}

\author{Lingyuan Kong}
\affiliation{%
  \institution{Kuaishou Technology}
  \city{Beijing}
  \country{China}
}
\email{konglingyuan03@kuaishou.com}

\author{Jiaqi Cui}
\affiliation{%
  \institution{Kuaishou Technology}
  \city{Beijing}
  \country{China}
}
\email{202521412456@uibe.edu.cn}

\author{Fanjiao Zeng}
\affiliation{%
  \institution{Kuaishou Technology}
  \city{Beijing}
  \country{China}
}
\email{zengfanjiao03@kuaishou.com}

\author{Congqi Wang}
\affiliation{%
  \institution{Kuaishou Technology}
  \city{Beijing}
  \country{China}
}
\email{wangcongqi@kuaishou.com}

\author{Yu Li}
\affiliation{%
  \institution{Kuaishou Technology}
  \city{Beijing}
  \country{China}
}
\email{liyu25@kuaishou.com}

\author{Yuan Cheng}
\affiliation{%
  \institution{Kuaishou Technology}
  \city{Beijing}
  \country{China}
}
\email{chengyuan03@kuaishou.com}

\author{Jingxin Liu}
\authornote{Corresponding author.}
%%\authornote{Both authors contributed equally to this research.}
%\email{webmaster@marysville-ohio.com}
\affiliation{%
  \institution{Kuaishou Technology}
  \city{Beijing}
  \country{China}
}
\email{liujingxin05@kuaishou.com}

\author{Xiaoshuang Chen}
\affiliation{%
  \institution{Kuaishou Technology}
  \city{Beijing}
  \country{China}
}
\email{chenxiaoshuang@kuaishou.com}

\author{Kaiqiao Zhan}
\affiliation{%
  \institution{Kuaishou Technology}
  \city{Beijing}
  \country{China}
}
\email{zhankaiqiao@kuaishou.com}


\renewcommand{\shortauthors}{Lingyuan Kong et al.}
%% No italics, no superscripts
%% Use footnote or author note to identify equal contribution and/or contact author info


% \setlength{\textfloatsep}{6pt plus 2pt minus 2pt}   % 浮动体与正文之间
% \setlength{\intextsep}{6pt plus 2pt minus 2pt}      % [h] 放置时的上下间距
% \setlength{\abovecaptionskip}{4pt}                  % 图与 caption 之间
% \setlength{\belowcaptionskip}{0pt}                  % caption 与下文之间

\begin{abstract}
Industrial recommender systems use cascaded stages with different candidate spaces, data distributions, and latency constraints. Each stage include multi-objective prediction and multi-task fusion (MTF) modules. Existing MTF methods typically design and optimize the fusion module independently for each stage, which can lead to inconsistent preferences across stages. Although cross-stage coordination methods have been proposed, they mainly transfer supervision to upstream scoring models rather than jointly optimizing the fusion modules.

We propose UniRec, a \textbf{Uni}fied Cross-stage \textbf{Rec}ommendation Fusion framework that jointly optimizes pre-ranking and ranking fusion agents. UniRec employs a coupled dual-agent architecture with partially shared embeddings and stage-specific fusion functions. It further introduces Dual-Axis Preference Optimization to combine stage-local preferences with downstream-to-upstream preference transfer. To enable stable and efficient cross-stage training, we address two practical challenges: reward-induced distribution drift and the computational cost of pairwise preference learning.

Extensive offline and online experiments on a production short-video system demonstrate consistent gains over single-stage fusion and cross-stage coordination baselines. In a nine-week online A/B test, UniRec improves total watch time by $0.675\%$ and app usage time by $0.616\%$, and active users by $0.189\%$, and has been fully deployed in production.
\end{abstract}


% <concept_id>10002951.10003317.10003347.10003350</concept_id>
% <concept_desc>Information systems~Recommender systems</concept_desc>

% <concept_id>10010147.10010257.10010258.10010261</concept_id>
% <concept_desc>Computing methodologies~Reinforcement learning</concept_desc>
% <concept_significance>300</concept_significance>

\ccsdesc[500]{Information systems~Recommender systems}
% \ccsdesc[300]{Computing methodologies~Reinforcement learning}

\keywords{Cascaded Recommender Systems, Cross-Stage Optimization, Multi-Task Fusion, Preference Optimization, Ranking Consistency}


\maketitle

\begin{figure}[htbp]
\centering
\includegraphics[width=0.9\columnwidth]{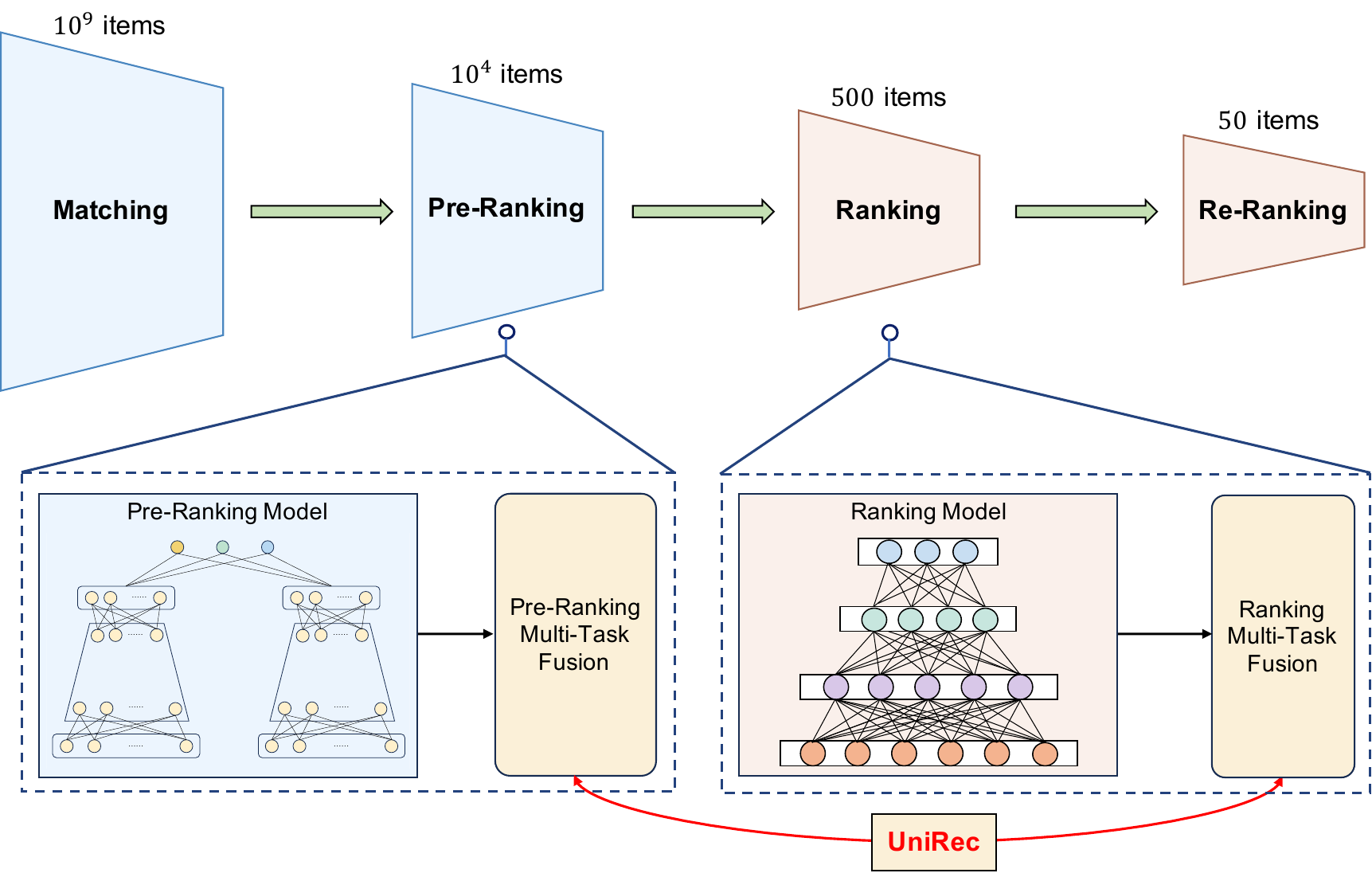}
\caption{Overview of the cascaded recommendation architecture and UniRec.}
\label{fig:overview}
% \vspace{-4pt}
\end{figure}

\section{Introduction}
\label{sec_intro}
Industrial recommender systems are commonly deployed as cascaded pipelines consisting of matching, pre-ranking, ranking, and re-ranking, balancing recommendation quality with strict online latency constraints \cite{wang2020cold}. The pre-ranking stage filters thousands of candidates under a tight budget, while downstream ranking performs more computationally intensive multi-objective optimization. As illustrated in Figure~\ref{fig:overview}, each stage contains two components: a multi-task prediction model estimates user feedback signals such as click, watch time, and like; and a multi-task fusion (MTF) module aggregates these predictions into a unified score determining which candidates survive the stage. Thus, the MTF module, rather than the prediction model, makes the final stage-wise selection decision.

\begin{figure}[htbp]
\centering
\includegraphics[width=0.9\columnwidth]{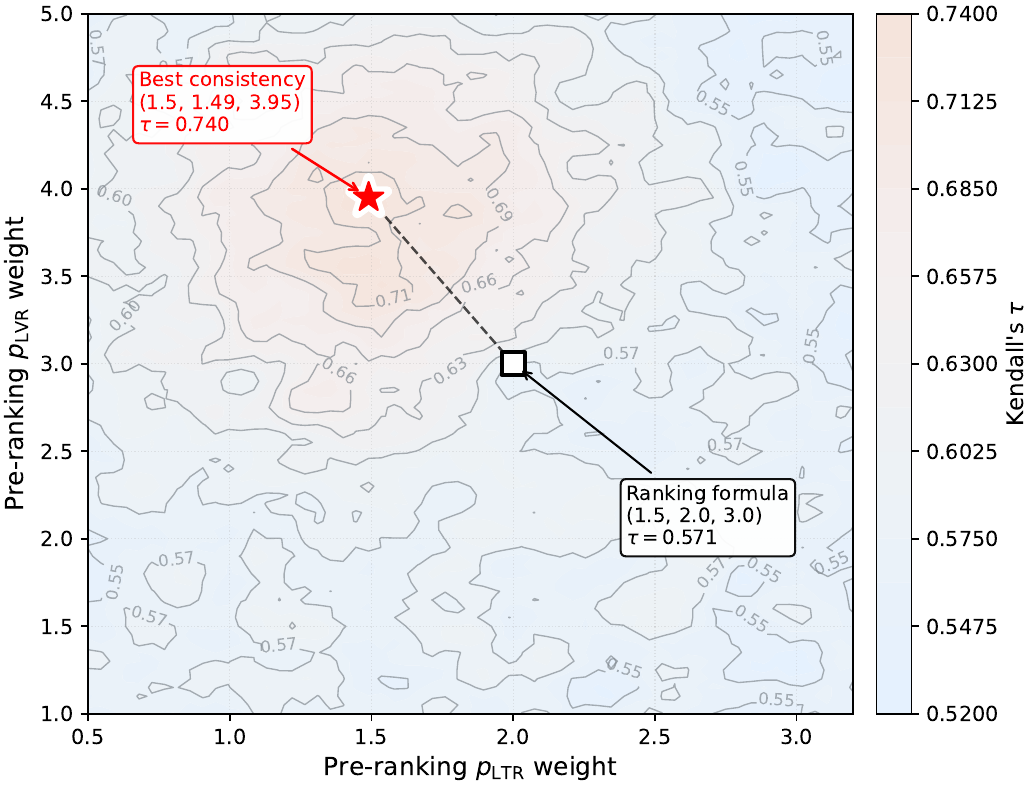}
\caption{Kendall's $\tau$ under different fusion weights.}
\label{fig:example}
% \vspace{-4pt}
\end{figure}

Most existing methods independently optimize the MTF modules of different stages, which can create cross-stage preference inconsistency. The key question is whether the ordering induced by pre-ranking remains consistent with the downstream ranking preferences. We quantify this consistency by measuring the rank correlation between pre-ranking and ranking fusion scores, using Kendall's $\tau$ over candidates that reach the ranking stage. Interestingly, directly aligning the fusion rules across stages does not necessarily yield the most consistent ranking. Using pXTR predictions collected from production traffic, we construct an illustrative ranking fusion rule $1.5 * \mathrm{pCTR}+2.0 * \mathrm{pLTR}+3.0 * \mathrm{pLVR}$, where the three terms denote the predicted click-through, like, and long-view rates. Fixing the pre-ranking pCTR coefficient at 1.5, we sweep its pLTR and pLVR coefficients over $[0,5]$ with a step size of 0.1, and measure Kendall’s $\tau$ between the resulting pre-ranking scores and the ranking scores over 10,000 randomly sampled requests. As shown in Figure~\ref{fig:example}, the coefficient combination achieving the highest consistency is not $(1.5,2.0,3.0)$, although $(1.5,2.0,3.0)$ is exactly the ranking formula. This counterintuitive result indicates that simply copying the downstream fusion rule cannot guarantee optimal cross-stage consistency. The two stages differ in candidate distributions and pXTR prediction quality, causing the same fusion function to induce different orderings across stages.

Therefore, \textbf{cross-stage MTF alignment is not a trivial problem: the pre-ranking fusion function must be adapted to its own stage while preserving preferences favored by downstream ranking.} Existing remedies mainly coordinate stages by transferring downstream supervision to upstream scoring models \cite{zhao2023copr, zhang2023rethinking, zhao2025hybrid}, but leave the MTF modules that make the actual selection decisions outside cross-stage optimization. Conversely, MTF research, ranging from hand-designed formulas and heuristic search \cite{rubinstein1999cross, movckus1974bayesian} to reinforcement learning \cite{zhang2022multi, zhang2024unex, cao2025xmtf} and end-to-end learning \cite{he2025end, xu2025umre}, mainly focuses on optimizing fusion within a single stage. These two lines of research therefore remain largely separate: cross-stage coordination improves upstream prediction or scoring models, while fusion optimization remains stage-local. 

This motivates us to develop a jointly optimized, unified MTF framework for multi-stage recommendation systems. However, jointly optimizing the two MTF modules introduces three challenges. \textbf{1. Heterogeneous cross-stage inputs.} The two stages operate on different candidate spaces and data distributions, with different prediction qualities and latency budgets. A unified framework must therefore couple the stages while preserving their stage-specific representations and fusion capacity. \textbf{2. Reward-induced distribution drift.} Joint optimization also amplifies the model's \emph{reward sensitivity}. When reward signals correlate with superficial item attributes, joint training incentivizes both stages to exploit such correlations, systematically biasing the recommendation distribution toward high–reward–but–not–preferred items rather than improving genuine preference alignment. 
 \textbf{3. Computational complexity of joint optimization.} Cross-stage joint optimization requires evaluating preference signals over the full pre-ranking candidate set, introducing non-trivial computational overhead that can be prohibitive for large-scale systems. Efficient joint optimization thus demands careful design to keep the added cost manageable.

To address these challenges, we propose UniRec, a \textbf{Uni}fied Cross-stage
\textbf{Rec}ommendation Fusion framework that jointly optimizes the pre-ranking
and ranking MTF modules. UniRec builds on a coupled dual-agent architecture with
partially shared embeddings and stage-specific fusion functions, so that both
stages are trained toward a single, consistent preference space. To further
enforce preference consistency, we introduce \emph{Dual-Axis Preference
Optimization}, which combines stage-local preference learning with explicit
downstream-to-upstream preference transfer.

Moreover, cross-stage optimization exposes two practical obstacles beyond
preference consistency. To control the reward-induced distribution drift, we
propose \emph{Attribute Group-Relative Regularization (AGRR)}, which constrains
both stages from uniformly exploiting reward-correlated attributes. To keep the
added cost manageable, we further propose \emph{Compact Pairwise Preference
Aggregation (CPPA)} with \emph{Confidence-Aware Pair Pruning (CAPP)}, which
reduces the number of evaluated candidate pairs. In a nine-week online A/B test
on a production short-video system, UniRec improves total watch time by
$0.675\%$ and active users by $0.189\%$. UniRec is fully deployed in production.

Our contributions are as follows:
\begin{itemize}
\item We formulate cross-stage multi-task fusion as a joint preference optimization problem and develop a coupled dual-agent architecture for jointly optimizing pre-ranking and ranking fusion modules.

\item We introduce Dual-Axis Preference Optimization, AGRR, and CPPA/CAPP to address cross-stage preference transfer, reward-induced distribution drift, and computational complexity.

\item We conduct extensive offline and online A/B experiments on a production short-video system, demonstrating consistent gains over cross-stage coordination and single-stage fusion baselines. UniRec is fully deployed in production.
\end{itemize}

\section{Related Work}
\subsection{Multi-task fusion and end-to-end ranking}

Industrial recommenders commonly formulate pre-ranking and ranking as multi-task problems, predicting multiple engagement objectives such as click-through rate, watch time, and like, and aggregating them into a unified score. Early systems relied on hand-designed ensemble formulas with coefficients tuned by heuristic methods such as the Cross-Entropy Method \cite{rubinstein1999cross} and Bayesian Optimization \cite{movckus1974bayesian}. Subsequent work learns fusion through reinforcement learning, including BatchRL-MTF \cite{zhang2022multi}, RLUR \cite{cai2023reinforcing}, and xMTF \cite{cao2025xmtf}. UNEX-RL \cite{zhang2024unex} further exploits the unidirectional dependency of cascaded systems through a multi-agent information chain, but its agents predict weights for fixed fusion formulas and do not exchange cross-stage gradients.

More recent work directly learns the fusion model through end-to-end optimization. EMER \cite{he2025end} adopts a transformer-based fusion architecture, while UMRE \cite{xu2025umre}, Pantheon \cite{cao2025pantheon}, and HarmonRank \cite{xia2026harmonrank} explore monotonic fusion, Pareto-based optimization \cite{sener2018multi, lin2019pareto}, and attention-based multi-objective alignment, respectively. Despite these advances, existing methods mainly optimize multi-task fusion within a single stage. Joint optimization of fusion modules across stages in industrial cascaded recommenders remains unexplored.

\subsection{Pre-ranking optimization and cross-stage coordination}

Recent studies improve pre-ranking through stronger models or coordination with downstream ranking. COLD \cite{wang2020cold} enables deeper pre-ranking models under latency constraints. COPR \cite{zhao2023copr} aligns pre-ranking with ranking through chunk-based sampling and $\Delta$NDCG-weighted loss, Rethink \cite{zhang2023rethinking} optimizes pre-ranking sets with the All-Scenario hit-rate and ASMOL, and HCCP \cite{zhao2025hybrid} transfers downstream feedback through hybrid sampling and margin-based InfoNCE \cite{oord2018representation}. These methods demonstrate the benefit of upstream supervision from downstream preferences.

However, existing coordination is primarily one-way: the ranking model acts as a teacher, while the two stages remain separately optimized. Moreover, they focus on the pre-ranking scoring model rather than jointly optimizing the multi-task fusion module that determines candidate selection. As illustrated in Figure~\ref{fig:overview}, even when upstream scores are better aligned with ranking, the fusion function itself may still produce decisions inconsistent with downstream preferences. UniRec addresses this gap by jointly optimizing the pre-ranking and ranking fusion module.

\section{Problem Formulation}

We consider multi-stage fusion optimization in cascaded recommender systems. For a user request, the cascade has $S$ stages. Stage $s$ scores each candidate in $\mathcal I^{(s-1)}$ with $s_i^{(s)}$ and retains $\mathcal I^{(s)}$:
\begin{equation}
s_i^{(s)} = f_s(\mathbf x_i^{(s)};\theta_s),
\quad
\mathcal I^{(s)} =
\operatorname*{TopK}_{i\in\mathcal I^{(s-1)}}(s_i^{(s)},K_s),
\quad s=1,\ldots,S
\end{equation}
Here $\mathcal I^{(0)}$ is the set retrieved by the matching module, with $\mathcal I^{(0)} \supseteq \mathcal I^{(1)} \supseteq \cdots \supseteq \mathcal I^{(S)}$. Stages may differ in candidate scale, input space, data distribution, and latency budget.We instantiate the cascade with pre-ranking ($p$) and ranking ($r$), its two critical stages. Their candidate sets are $\mathcal I^c=\mathcal I^{(0)}$ from matching, $\mathcal I^p=\mathcal I^{(1)}$ from pre-ranking, and $\mathcal I^r=\mathcal I^{(2)}$ from ranking.

Both stages are built upon multi-task learning models \cite{ma2018modeling,tang2020progressive}, where the backbone jointly predicts multiple task-specific pXTRs for each candidate $i$, denoted as $\mathbf p_i = [p_i^{(1)},\cdots,p_i^{(M)}]$, with $M$ being the number of prediction tasks. At each stage, a fusion module $f_t(\cdot;\theta_t)$, $t\in\{p,r\}$, produces a unified score from input $\mathbf x_i^t$, which combines context, user, and item features with the corresponding pXTRs. The stage scores are
\begin{equation}
s_i^{p} = f_p(\mathbf x_i^{p};\theta_p),
\qquad
s_i^{r} = f_r(\mathbf x_i^{r};\theta_r)
\end{equation}
and each stage retains its top-scored candidates:
\begin{equation}
\mathcal I^{p} = \operatorname*{TopK}_{i \in \mathcal I^{c}}(s_i^{p},K_p),
\qquad
\mathcal I^{r} = \operatorname*{TopK}_{i \in \mathcal I^{p}}(s_i^{r},K_r)
\label{eq:topk}
\end{equation}

Existing recommendation systems optimize the two fusion modules separately, each with its own stage-local objective:
\begin{equation}
\min_{\theta_p}\ \mathcal L_p(\theta_p),
\qquad
\min_{\theta_r}\ \mathcal L_r(\theta_r)
\label{eq:separate}
\end{equation}
Following single-stage fusion methods such as EMER \cite{he2025end}, each stage-local objective can combine pairwise losses from prior pXTR signals, which provide task-specific candidate orderings, and posterior user feedback, which provides preferences over exposed items. Although EMER focuses on ranking, this prior-posterior supervision can be applied to either stage, so we leave $\mathcal L_p$ and $\mathcal L_r$ abstract here.

However, the two objectives cannot be optimized independently without considering their interaction. The pre-ranking and ranking stages operate on different candidate spaces and data distributions under different computational budgets, and their pXTRs can have different prediction quality. Even when the underlying objectives are the same, their optimal fusion functions may differ. We therefore formulate cross-stage fusion as the joint optimization problem:
\begin{equation}
\min_{\theta_p,\ \theta_r}\ \mathcal L(\theta_p,\theta_r)
\label{eq:joint_obj}
\end{equation}
Joint preference optimization in Eq.~\eqref{eq:joint_obj} faces three challenges: coupling stages with different candidate distributions, prediction qualities, and latency budgets while preserving their stage-specific representations and fusion capacity; mitigating reward-induced distribution drift without sacrificing within-group preference ordering; and controlling the quadratic pairwise cost, especially over the larger pre-ranking candidate set $\mathcal I^c$.

\begin{figure*}[htbp]
\centering
\includegraphics[width=0.95\textwidth]{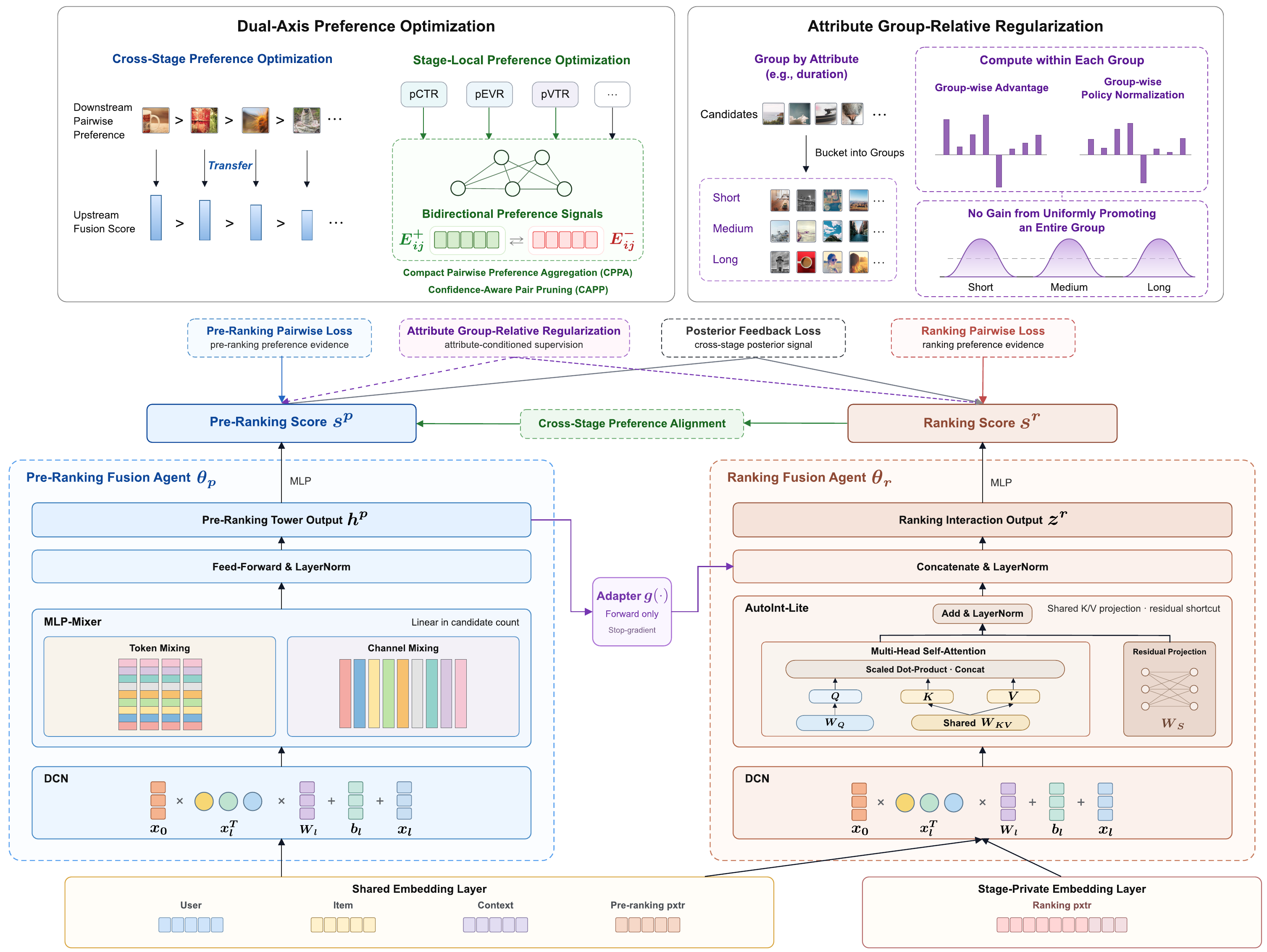}
\caption{The overall framework of UniRec. }
% \vspace{-4mm}
\label{fig:framework}
\end{figure*}

\section{Method}
\label{sec:method}
Figure~\ref{fig:framework} illustrates the overall architecture of UniRec. UniRec places a fusion agent at each stage and connects them through a single differentiable graph, enabling joint optimization rather than independent stage-wise training. The fusion agents operate solely on the outputs of existing stage-specific prediction models and remain decoupled from them. Sec.~\ref{sec:arch} and \ref{sec:joint} unify the heterogeneous inputs of different stages for joint optimization. Sec.~\ref{sec:grpo} introduces attribute group-relative regularization (AGRR) to mitigate reward-induced distribution drift, while Sec.~\ref{sec:align} reduces the computational complexity of multi-stage pairwise preference optimization through compact pairwise preference aggregation (CPPA) and confidence-aware pair pruning (CAPP).

\subsection{Coupled Dual-Agent Architecture}
\label{sec:arch}

UniRec places a fusion agent at each stage and couples the two into a single optimization problem, rather than optimizing them independently. The architecture reconciles their heterogeneous candidate spaces, data distributions, and prediction qualities while preserving stage-specific fusion capacity.

\subsubsection{Partially shared embedding space.}
Both agents share a common embedding layer encoding features available to both stages, including user, item, and context features, as well as the pre-ranking task predictions $\mathbf p_i^{p}$. Ranking-specific predictions $\mathbf p_i^{r}$ are unavailable to pre-ranking and are therefore included only in the ranking-specific embedding. Let $\mathbf e^{\mathrm{sh}}_i$ denote the shared embedding and $\mathbf e^{r}_i$ the ranking-specific embedding. The two agents receive
$\mathbf x_i^{p} = \mathbf e^{\mathrm{sh}}_i$ and
$\mathbf x_i^{r} = [\mathbf e^{\mathrm{sh}}_i, \mathbf e^{r}_i]$.

The shared embedding makes the two stages part of a single computation graph: $\mathcal L_{\mathrm{pre}}$ and $\mathcal L_{\mathrm{rank}}$ update the same parameters $\theta_{\mathrm{sh}}$, allowing gradients from either stage to reach the shared representation. Thus, cross-stage coupling is symmetric despite the asymmetric input spaces, unlike approaches where a frozen downstream model only supervises the upstream stage.

\subsubsection{Stage-specific fusion towers.}
Each fusion agent has an independent tower, $f_p$ and $f_r$, whose architecture can be customized to its computational budget. Both towers use Deep Cross Network (DCN) \cite{wang2017deep} for explicit feature crosses, followed by a stage-specific block. The pre-ranking agent uses a lightweight MLP-Mixer \cite{tolstikhin2021mlp}, whose cost scales linearly with the number of candidates. The ranking agent uses AutoInt-Lite, a reduced AutoInt \cite{song2019autoint} block with shared key/value projections, a single interaction layer, a small number of heads, and a linear shortcut. This asymmetry follows the stage-wise computational constraints: pre-ranking handles an order of magnitude more candidates under a tighter latency budget, while ranking operates on a smaller candidate set and can afford lightweight attention to model candidate dependencies.

\subsubsection{Adapter network.}
To incorporate the upstream decision into ranking, an adapter $g(\cdot)$ projects the pre-ranking representation $\mathbf h_i^{p}$ to the ranking tower's dimensionality. Its output is concatenated with the AutoInt-Lite representation $\mathbf z_i^{r}$ before obtaining $s_i^r$:
\begin{equation}
s_i^{r} = \operatorname{MLP}\big(\big[\mathbf z_i^{r},\ g\big(\operatorname{sg}[\mathbf h_i^{p}]\big)\big]\big)
\end{equation}
where $\operatorname{sg}[\cdot]$ denotes stop-gradient. Detaching $\mathbf h_i^{p}$ prevents the ranking objective from reshaping the pre-ranking representation toward downstream scoring. Thus, cross-stage gradient coupling is restricted to the shared embedding space, while the adapter provides the ranking agent with the upstream decision as a feature.

\subsection{Two-Stage Joint Optimization}
\label{sec:joint}

A fusion agent is supervised by \emph{preferences} rather than absolute targets, with scores trained to preserve desired candidate orderings. We organize joint optimization along two complementary axes, termed \emph{Dual-Axis Preference Optimization}. \emph{Stage-locally}, each agent receives prior pXTR preferences and posterior user feedback. \emph{Cross-stage}, downstream preferences are transferred from ranking to pre-ranking, coordinating the two stages while preserving their stage-specific fusion functions. We define the score margins as
\begin{equation}
\delta_{ij}^{p} \triangleq s_i^{p}-s_j^{p},
\qquad
\delta_{ij}^{r} \triangleq s_i^{r}-s_j^{r}
\end{equation}

\subsubsection{Stage-Local Preference Optimization}
For prior supervision, each pXTR induces pairwise preferences over candidates. Let $M_p$ and $M_r$ denote the numbers of pXTR objectives in the pre-ranking and ranking stages. For objective $m$, the preference label is:
\begin{equation}
y_{ij}^{(m)}=\operatorname{sign}\left(p_i^{(m)}-p_j^{(m)}\right)
\end{equation}
with tied pairs excluded. The pre-ranking agent receives two sources of prior supervision: its own pXTRs over the full candidate set $\mathcal I^{c}$ and ranking pXTRs over the subset $\mathcal I^{p}$ that reaches ranking. The prior loss of the pre-ranking agent is defined as:
\begin{equation}
\begin{aligned}
\mathcal{L}_{prior}^{pre} =& \sum_{m=1}^{M_p} w_m
\sum_{i,j\in\mathcal I^{c}}
\operatorname{Softplus}
\big(-y_{ij}^{p,m}\delta_{ij}^{p}\big) \\
&+
\sum_{m=1}^{M_r} w_m
\sum_{i,j\in\mathcal I^{p}}
\operatorname{Softplus}
\big(-y_{ij}^{r,m}\delta_{ij}^{p}\big)
\end{aligned}
\label{eq:prior_pre}
\end{equation}
where $y_{ij}^{p,m}$ and $y_{ij}^{r,m}$ are the pairwise labels induced by the pre-ranking and ranking pXTRs. Thus, ranking pXTRs provide additional training-time supervision to pre-ranking on $\mathcal I^{p}$ without being used as its inputs. The ranking agent is supervised only by ranking pXTR preferences over $\mathcal I^{p}$:
\begin{equation}
\mathcal{L}_{prior}^{rank} = \sum_{m=1}^{M_r} w_m \sum_{i,j\in\mathcal I^{p}}\operatorname{Softplus} \big(-y_{ij}^{r,m}\delta_{ij}^{r}\big)
\label{eq:prior_rank}
\end{equation}

Exposed items additionally provide posterior user feedback. For the exposure item set $\mathcal I^{e}$, let
$y_{ij}^{e}=\operatorname{sign}(r_i^{e}-r_j^{e})$,
where $r_i^{e}$ denotes the observed feedback or reward. The posterior preference loss for either stage is
\begin{equation}
\mathcal{L}_{feedback} = \sum_{i,j\in\mathcal I^{e}} \operatorname{Softplus}\big(-y_{ij}^{e}(s_i-s_j)\big)
\label{eq}
\end{equation}
Since only a small fraction of candidates are exposed, this posterior signal serves as auxiliary supervision.

\subsubsection{Cross-Stage Preference Optimization}
\label{sec:vertical}

Stage-local optimization does not directly account for downstream ordering among candidates that survive pre-ranking. We therefore transfer the ranking preference
$y_{ij}^{r}=\operatorname{sign}(\delta_{ij}^{r})$
to the pre-ranking score:
\begin{equation}
\mathcal{L}_{align} = \sum_{i,j\in\mathcal I^{p}}\operatorname{Softplus}\left(-\operatorname{sg}[y_{ij}^{r}]\delta_{ij}^{p}\right)
\label{eq:align}
\end{equation}
where $\operatorname{sg}[\cdot]$ denotes stop-gradient. Pairs are drawn from $\mathcal I^{p}$ because filtered candidates have no downstream score. Rather than matching the ranking score or fusion function, this objective transfers only downstream ordering preferences, allowing the two stages to retain different fusion functions while encouraging pre-ranking to preserve downstream-preferred items.

\subsection{Attribute Group-Relative Regularization}
\label{sec:grpo}

Joint optimization makes the cascade more sensitive to reward design. This sensitivity is particularly problematic in recommendation systems, where rewards often exhibit inherent attribute biases and can therefore be exploited through reward hacking. In short-video recommendation, for example, longer videos may receive higher watch-time rewards simply because of their duration, rather than because users prefer them. An agent can then reduce the pairwise loss by promoting long videos as a group, without improving the ordering of videos within that group. Joint optimization may propagate and amplify this bias across stages. To mitigate it, we introduce a group-relative regularizer that compares items within similar attribute ranges, following the principle of GRPO \cite{shao2024deepseekmath}.

\textbf{Reward from preference signals.}
Rather than introducing a separate reward model, we derive the item-level reward directly from the pXTR preferences used by each fusion agent. The ranking agent uses only the ranking pXTR preferences over $\mathcal I^{p}$, while the pre-ranking agent combines its own pre-ranking pXTR preferences and, for candidates in $\mathcal I^{p}$, the additional ranking pXTR preferences. The reward of item $i$ is then defined as
\begin{equation}
r_i =
\frac{\sum_{j}\big(E_{ij}^{+}-E_{ij}^{-}\big)}
{\sum_{j}\big(E_{ij}^{+}+E_{ij}^{-} +\epsilon\big)}
\in[-1,1]
\label{eq:cppa_reward}
\end{equation}
where $\epsilon>0$ prevents division by zero, $E_{ij}^{+}$ and $E_{ij}^{-}$ for each item pair $(i,j)$ are detailed in Sec. \ref{sec:cppa}. Using the same objectives and weights as the main supervision keeps the reward aligned with the original preference signals.

\textbf{Group-relative advantage.}
Candidates are partitioned into $B$ quantile buckets ${\mathcal G}_1,\ldots,\mathcal G_B$ by the attribute $a_i$ (e.g., video duration). For $i\in\mathcal G_g$, its advantage is normalized within its own group:
\begin{equation}
A_i =
\operatorname{clip}\left(
\frac{r_i-\mu_g}{\max(\sigma_g,\sigma_{\min})},
-A_{\max},A_{\max}\right),
\quad i\in\mathcal G_g
\label{eq:advantage}
\end{equation}
where $\mu_g$ and $\sigma_g$ are the mean and standard deviation of rewards in $\mathcal G_g$; buckets with fewer than $n_{\min}$ candidates are skipped. The AGRR objective is invariant to a uniform score shift within an attribute group and therefore does not reinforce such group-level shifts.

\textbf{Within-group policy regularization.}
The advantages define a target distribution $q(i\mid g)\propto\exp(A_i/\tau)$, while the model induces $\pi_\theta(i\mid g)\propto\exp(s_i)$, with both normalized over $\mathcal G_g$. We optimize their KL divergence:
\begin{equation}
\mathcal L_{\mathrm{AGRR}} = \frac{1}{|\mathcal B|} \sum_{g\in\mathcal B}\operatorname{KL}\big(q(\cdot\mid g)\| \pi_\theta(\cdot\mid g)\big)
\label{eq:grpo}
\end{equation}
where $\mathcal B$ contains buckets passing the count threshold. We use KL instead of the advantage-weighted likelihood $-\sum_iA_i\log\pi_\theta$, which can become unbounded under score scaling. In contrast, the KL gradient with respect to the group logits is $\pi_\theta(i\mid g)-q(i\mid g)$ and vanishes as the policy approaches the target.

\subsection{Overall Joint Objective}

Combining the two axes, the fusion agents are jointly optimized with stage-local preferences, cross-stage alignment, and AGRR:
\begin{equation}
\begin{aligned}
\mathcal{L}_{pre} &= \mathcal{L}_{prior}^{p} +\lambda_{\mathrm{fb}}^{pre}\mathcal{L}_{feedback}^{pre} +\lambda_{\mathrm{align}}\mathcal{L}_{align} +\lambda_{\mathrm{AGRR}}^{pre}\mathcal{L}_{AGRR}^{pre} \\
\mathcal{L}_{rank} &=
\mathcal{L}_{prior}^{r} +\lambda_{\mathrm{fb}}\mathcal{L}_{feedback}^{rank}
+\lambda_{\mathrm{AGRR}}\mathcal{L}_{AGRR}^{rank} \\
\mathcal{L}_{\mathrm{UniRec}} &= \mathcal{L}_{pre} + \mathcal{L}_{rank}
\end{aligned}
\label{eq:total}
\end{equation}
The alignment term appears only for pre-ranking because it transfers downstream preferences upstream, while the shared representation allows both objectives to update the common parameters.

As shown in Fig.~\ref{fig:overview}, the two fusion agents serve as the multi-task fusion modules for pre-ranking and ranking, following a centralized-training, decentralized-execution paradigm \cite{lowe2017multi}. During training, they are jointly optimized and coupled through the shared representation. At deployment, they are exported as separate graphs with the shared embedding table, so cross-stage coupling exists only during training. Each agent then operates as a stage-local fusion model without additional online latency.

\subsection{Computational Complexity Optimization}
\label{sec:align}

The prior preference losses in Eq.~(\ref{eq:prior_pre})--Eq.~(\ref{eq:prior_rank}) generate pairwise supervision for multiple pXTR objectives at both stages. In two-stage joint optimization, this cost is amplified by the large candidate set of pre-ranking and the additional preference signals transferred from ranking pXTRs. For $M$ applicable pXTR objectives and $N$ candidates, directly evaluating the pairwise losses requires $\mathcal{O}(MN^2)$ Softplus terms. We therefore optimize the pairwise supervision through CPPA and CAPP strategies.

\subsubsection{Compact Pairwise Preference Aggregation}
\label{sec:cppa}

For a fixed candidate pair $(i,j)$, all objective-specific losses share the same score margin $\delta_{ij}$ and differ only in preference direction. We therefore aggregate objectives with the same direction before applying the nonlinear loss. Define
\begin{equation}
E_{ij}^{+} = \sum_{m=1}^{M} w_m \mathbb{I}\big(y_{ij}^{(m)}=+1\big),
\qquad
E_{ij}^{-} = \sum_{m=1}^{M} w_m \mathbb{I}\big(y_{ij}^{(m)}=-1\big)
\label{eq:evidence}
\end{equation}
where $E_{ij}^{+}$ and $E_{ij}^{-}$ denote the weighted support for ranking $i$ ahead of $j$ and vice versa. Both directions are retained, thereby preserving disagreement among objectives. CPPA then rewrites the pairwise objective as
\begin{equation}
\begin{aligned}
\mathcal{L}_{agg}^{+} &= \sum_{i,j}E_{ij}^{+}\operatorname{Softplus}\big(-\delta_{ij}\big), \;
\mathcal{L}_{agg}^{-} = \sum_{i,j}E_{ij}^{-}\operatorname{Softplus}\big(\delta_{ij}\big)
\end{aligned}
\label{eq:agg}
\end{equation}
Then the final CPPA loss is defined as $\mathcal{L}_{CPPA} = \mathcal{L}_{agg}^{+} + \mathcal{L}_{agg}^{-}$.

CPPA is algebraically equivalent to the original objective, as the $M$ objective-specific terms for each pair are grouped by preference direction before applying Softplus. It thus preserves all pairwise preference information while reducing Softplus evaluations from $\mathcal{O}(MN^2)$ to $\mathcal{O}(N^2)$. The same reorganization applies to both stages, including the ranking-pXTR supervision transferred to pre-ranking.

\subsubsection{Confidence-Aware Pair Pruning}

Although CPPA removes the objective-dependent factor $M$, evaluating all $\mathcal{O}(N^2)$ pairs can remain expensive for large candidate sets. We therefore further prune pairs according to the consistency of their aggregated preference evidence. For each pair $(i,j)$, we define its confidence as the absolute difference between the two directional supports: $c_{ij} = \left|E_{ij}^{+}-E_{ij}^{-}\right|$. A larger $c_{ij}$ indicates stronger preference evidence in one direction and hence a more informative pair. For each candidate $i$, we sort the other candidates by $c_{ij}$ and retain the top-$K$ pairs:
\begin{equation}
\mathcal{P}_{K} =
\left\{(i,j)\mid j\in\operatorname{TopK}_{j}(c_{ij})
\right\}
\label{eq:pair_pruning}
\end{equation}
The resulting pair set contains at most $\mathcal{O}(KN)$ pairs, and only these pairs are used for pairwise loss evaluation. Thus, CAPP reduces the number of differentiable pairwise loss evaluations from $O(N^2)$ to $O(KN)$. Together with CPPA, which removes the objective-dependent factor $M$ from loss evaluation, CAPP substantially reduces the computational cost of pairwise preference optimization.

\section{Experiments}

We evaluate UniRec through four research questions:
\begin{itemize}
\item \textbf{RQ1:} Does UniRec outperform baseline methods in ranking quality and cross-stage consistency?
\item \textbf{RQ2:} How much does each major component of UniRec contribute to its performance and efficiency?
\item \textbf{RQ3:} How sensitive is UniRec to the loss weights?
\item \textbf{RQ4:} Does UniRec improve user engagement in a large-scale online recommendation system?
\end{itemize}

\begin{table*}[t]
\centering
\caption{Comparison of ranking performance and cross-stage consistency.}
\label{tab:main_results}
\begin{tabular}{lcccccccccc}
\toprule
Method & rank\_GAUC
& \multicolumn{3}{c}{Ranking NDCG}
& \multicolumn{3}{c}{Pre-ranking NDCG}
& \multicolumn{3}{c}{Cross-stage Consistency} \\
\cmidrule(lr){3-5}
\cmidrule(lr){6-8}
\cmidrule(lr){9-11}
& & @1 & @3 & @5
& @1 & @3 & @5
& ASH & Kendall's $\tau$ & Spearman's $\rho$ \\
\midrule
Weighted-Sum
& 0.6111 & 0.3921 & 0.4999 & 0.5992
& 0.3891 & 0.4966 & 0.5966
& 0.9809 & 0.4844 & 0.6450 \\

EMER
& 0.6158 & 0.3837 & 0.4941 & 0.5977
& 0.3885 & 0.4967 & 0.5965
& 0.8566 & -0.0444 & -0.0542 \\

UMRE
& 0.6096 & 0.3934 & 0.5008 & 0.6001
& 0.3885 & 0.4967 & 0.5965
& 0.9728 & 0.4545 & 0.6086 \\

COPR
& 0.6089 & 0.3916 & 0.4996 & 0.5990
& 0.3952 & 0.5011 & 0.5999
& 0.9913 & 0.5150 & 0.6802 \\

\midrule
\textbf{UniRec (Ours)}
& \textbf{0.6330} & \textbf{0.3975} & \textbf{0.5099} & \textbf{0.6090}
& \textbf{0.3970} & \textbf{0.5075} & \textbf{0.6066}
& \textbf{0.9928} & \textbf{0.6017} & \textbf{0.7644} \\
\bottomrule
\end{tabular}
\end{table*}

\subsection{Experimental Setup}

\subsubsection{RecFlow: Public Cascaded Benchmark}
RecFlow\footnote{\url{https://github.com/RecFlow-ICLR/RecFlow}} is collected from Kuaishou's multi-stage short-video recommendation system and contains 37 days of logs from January 13 to February 18, 2024. It covers about 42K users, 9.3M requests, 38M exposure samples, and 1.9B stage-level records. The system has six stages, with candidate sizes decreasing from $8000 \to 3000 \to 500 \to 100 \to 10 \to 6$. RecFlow records the candidates removed at each stage, allowing us to reconstruct the nested candidate sets $\mathcal{I}^{c} \supset \mathcal{I}^{p} \supset \mathcal{I}^{r}$ in Eq.~\ref{eq:topk} and the shared candidate set $\mathcal{I}^{p}$ required by $\mathcal{L}_{align}$ in Eq.~\ref{eq:align}. 

We use eight days of exposure logs from February 1 to February 8. Each sample contains cross-stage candidate information and a 50-step behavior sequence. Each item has six binary labels and a \emph{playing\_time} label. Following \cite{covington2016deep}, we process \emph{playing\_time} with weighted logistic regression, yielding $M{=}7$ prediction inputs for the fusion model, with positive rates ranging from 0.2\% to 38.2\%.

\subsubsection{Industrial Production Dataset}
We also evaluate UniRec on data from Kuaishou's online recommendation system, which serves more than 100M daily active users. For each request, the pre-ranking stage selects candidates from about 500 items, which are further reduced to about 10 exposed items by subsequent stages. We sample about 100K users for offline evaluation, obtaining more than 1M request-level samples. The fusion model handles dozens of prediction objectives, making the $\mathcal{O}(MN^2)$ complexity of the per-objective formulation in Eq.~\ref{eq:prior_pre} expensive and providing a suitable setting for evaluating CPPA. The logs are split chronologically.

\subsubsection{Baselines}
We compare UniRec with four representative fusion methods, covering weighted fusion, single-stage learnable fusion, and cross-stage coordination. Only Weighted-Sum and UniRec use trainable fusion modules at both stages; EMER and UMRE operate at ranking, while COPR operates at pre-ranking, with the other stage using a fixed-weight sum of pXTRs.

\begin{itemize}
\item \textbf{WeightedSum}~\cite{rubinstein1999cross,movckus1974bayesian}
combines multiple prediction scores using learned weights and is trained at both stages with the same posterior feedback loss as UniRec. It uses neither item representations nor list context.

\item \textbf{EMER}~\cite{he2025end} uses a Transformer-based list encoder trained with an order-preserving loss and posterior feedback at the ranking stage.

\item \textbf{UMRE}~\cite{xu2025umre} applies a monotonic transformation to prediction scores at the ranking stage, with intent-aware weights for each item.

\item \textbf{COPR}~\cite{zhao2023copr} adjusts pre-ranking scores to match the downstream ordering through a pairwise loss weighted by $\Delta$NDCG, updating only the upstream model.
\end{itemize}

\subsubsection{Implementation Details}
RecFlow does not provide the pXTRs of its pre-ranking and ranking models, so we train two $M{=}7$-head backbones to generate $\mathbf{p}_i^p$ and $\mathbf{p}_i^r$, using DSSM and DIN, respectively, to match the different capacity requirements of the two stages. The backbones are trained on February 1--4 and then frozen; the fusion models are trained on February 5--7 and evaluated on February 8. All methods use the same data splits, cached pXTRs, candidate sets, and posterior feedback to isolate the effect of the fusion module.

We implement all methods in PyTorch and optimize them with Adam using a learning rate of $10^{-3}$, a batch size of 256, and a shared hidden dimension of 64. Baseline-specific hyperparameters follow their original papers or official implementations. All methods are tuned under the same validation protocol and search budget. The loss weights of UniRec are reported in Sec.~\ref{sec:sensitivity}.

\begin{figure*}[htbp]
\centering
\includegraphics[width=1.0\textwidth]{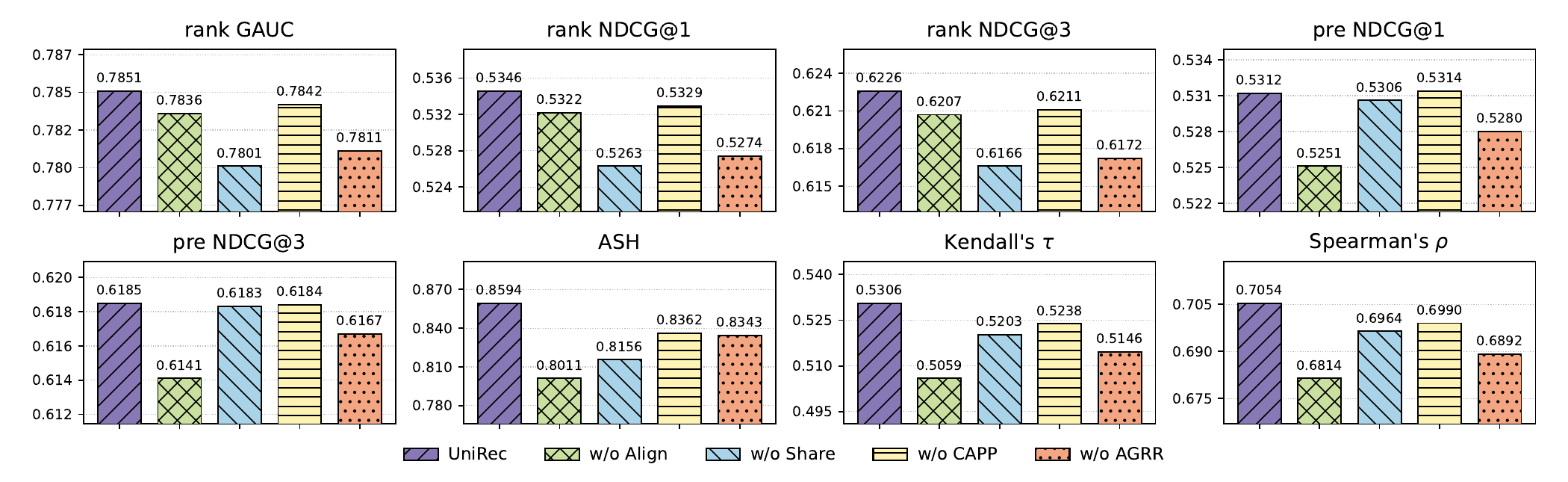}
\caption{Ablation study on the industrial production stream.}
\label{fig:ablation}
% \vspace{-4pt}
\end{figure*}

\subsection{Offline Evaluation (RQ1)}
\label{sec:offline}

Following RecFlow~\cite{liu2024recflow}, we evaluate candidates on a per-request basis. NDCG@$K$ is computed for both $s^r_i$ and $s^p_i$ over the exposed set $\mathcal{I}^{e}$ of each request, with $K \in \left\{1,3,5\right\}$ and $g_i$ denoting the ground-truth label. Restricting the cutoffs to exposed items keeps the metric focused on candidates for which feedback is observed. We report rank\_GAUC, which computes AUC within each request using $\mathbb{I}[g_i>0]$ as the binary label and then averages across requests, avoiding domination by cross-request score calibration. For cross-stage consistency, we report ASH~\cite{zhang2023rethinking}, the fraction of ranking-stage Top-$K_r$ items retained by pre-ranking with $K_r{=}10$ and $K_p{=}50$, together with Kendall's $\tau$ and Spearman's $\rho$, which measure rank correlation between $s^p_i$ and $s^r_i$ within each request.

\textbf{Ranking quality. }
Table~\ref{tab:main_results} reports the results on RecFlow. UniRec achieves the best rank\_GAUC and NDCG at all six cutoffs, improving rank\_GAUC from 0.6158 to 0.6330 over the strongest baseline. The gain is consistent across both stages: UniRec achieves a pre-ranking NDCG@3 of 0.5075, exceeding both the fixed-weight reference (0.4967) and COPR (0.5011), which explicitly optimizes the upstream score toward the downstream ordering. Joint optimization of both fusion modules therefore improves the upstream decision beyond one-way downstream supervision.

\textbf{Cross-stage consistency. }
EMER and UMRE have identical pre-ranking NDCG because neither trains an upstream fusion module; both use the fixed-weight sum of pXTRs. For these single-stage methods, the consistency metrics therefore measure how far the trained stage deviates from this common reference. EMER scores items using candidate context and preserves pXTR preferences only through a soft loss rather than by construction. Without explicit cross-stage alignment, it achieves competitive ranking quality but the lowest consistency. Training both stages alone is also insufficient: Weighted-Sum, the only other method trained at both stages, achieves $\tau=0.4844$, below COPR, which trains only the pre-ranking stage but explicitly aligns it with the downstream ordering. UniRec combines joint optimization with explicit alignment and achieves the highest values on all three consistency metrics, with $\tau=0.6017$ and $\rho=0.7644$, compared with 0.5150 and 0.6802 for COPR. ASH differentiates the methods less clearly because both UniRec (0.9928) and COPR (0.9913) are near saturation at $K_p{=}50$. In contrast, the rank correlations consider the full orderings and show that UniRec produces closer agreement between the two stages than one-way downstream-to-upstream alignment.

\subsection{Ablation Study (RQ2)}
\label{sec:ablation}
\begin{figure}[htbp]
\centering
\includegraphics[width=0.9\columnwidth]{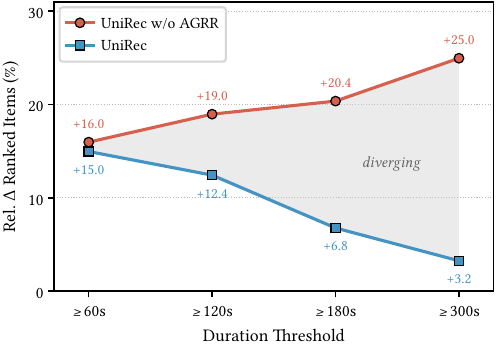}
\caption{Long-tail amplification with and without AGRR.}
\label{fig:app_duration_threshold}
% \vspace{-10pt}
\end{figure}

We answer RQ2 on the industrial production dataset. Each variant removes one component from the full model, with all other settings unchanged. \textbf{The components affect different metrics.}

As shown in Fig.~\ref{fig:ablation}, the full model is optimal on both the ranking-quality and cross-stage consistency metrics. \textbf{w/o Share} causes the largest ranking degradation, with rank\_GAUC dropping from 0.7851 to 0.7801 and rank\_NDCG@1 from 0.5346 to 0.5263, indicating that sharing the bottom embedding between the pre-ranking and ranking models is the primary driver of ranking quality. \textbf{w/o AGRR} reduces rank\_GAUC by 0.0040 and ASH by 0.0251, showing that AGRR contributes to both ranking performance and cross-stage consistency. \textbf{w/o CAPP} yields a marginal ranking drop ($-$0.0009 rank\_GAUC) but reduces ASH by 0.0232, suggesting that confidence-aware pair pruning primarily benefits cross-stage consistency rather than ranking accuracy.

\textbf{Alignment improves only the upstream stage.} \textbf{w/o Align} leaves ranking metrics largely unchanged (rank\_GAUC: $0.7851 \to 0.7836$), consistent with the design of $\mathcal{L}_{\mathrm{align}}$: the ranking score serves as a detached teacher signal and sends no gradient to ranking-specific parameters. Its effect is confined to the upstream stage and cross-stage consistency, where removing it reduces ASH by 0.0583 and Kendall's $\tau$ by 0.0247. This supports the design in Sec.~\ref{sec:vertical}, which transfers downstream preferences upstream without compromising ranking quality.

As shown in Section \ref{sec_intro}, joint optimization amplifies reward sensitivity. In short-video recommendation system, longer videos naturally accumulate more positive watch-time evidence, driving both stages to promote high-duration items as a group rather than improving within-group ordering. Fig.~\ref{fig:app_duration_threshold} confirms that without AGRR, the system over-promotes long videos in the final ranking output ($\geq$300s: +25.0\% relative to base, vs.\ +3.2\% for full UniRec), while AGRR suppresses this drift and aligns the two stages.The full distribution comparison can be found in Appendix Figure \ref{agrr_duration}.

\begin{table}[htbp]
\centering
\caption{Training efficiency comparison between UniRec (CPPA) and UniRec-NoCPPA.}
\label{tab:efficiency}
\begin{tabular}{lcc}
\toprule
Component & UniRec (CPPA) & UniRec-NoCPPA \\
\midrule
loss\_fn (rank)   & $47.17 \pm 15.92$ ms  & $241.62 \pm 43.02$ ms \\
loss\_fn (pre)    & $52.16 \pm 25.15$ ms  & $241.39 \pm 44.93$ ms \\
compute\_loss     & $187.35 \pm 41.15$ ms & $538.72 \pm 56.08$ ms \\
train\_step       & $323.64 \pm 70.29$ ms & $793.19 \pm 113.08$ ms \\
\midrule
Throughput (req/s) & \textbf{49}          & 20 \\
\bottomrule
\end{tabular}
\end{table}

Finally, since CPPA is mathematically equivalent to the per-objective formulation and affects only training efficiency, we omit \textbf{w/o CPPA} from the ablation. Table~\ref{tab:efficiency} shows that compared with naive pairwise loss, CPPA reduces loss computation by 5$\times$ and improves throughput by 2.5$\times$ (49 vs.\ 20 req/s), and Appendix Fig.~\ref{app_cppa_scale} shows that CPPA consistently reduces computation time across different numbers of objectives. Moreover, the computational cost results of CAPP strategy can be found in Appendix Fig.~\ref{app_diff_k}.

\begin{figure}[htbp]
\centering
\includegraphics[width=0.9\columnwidth]{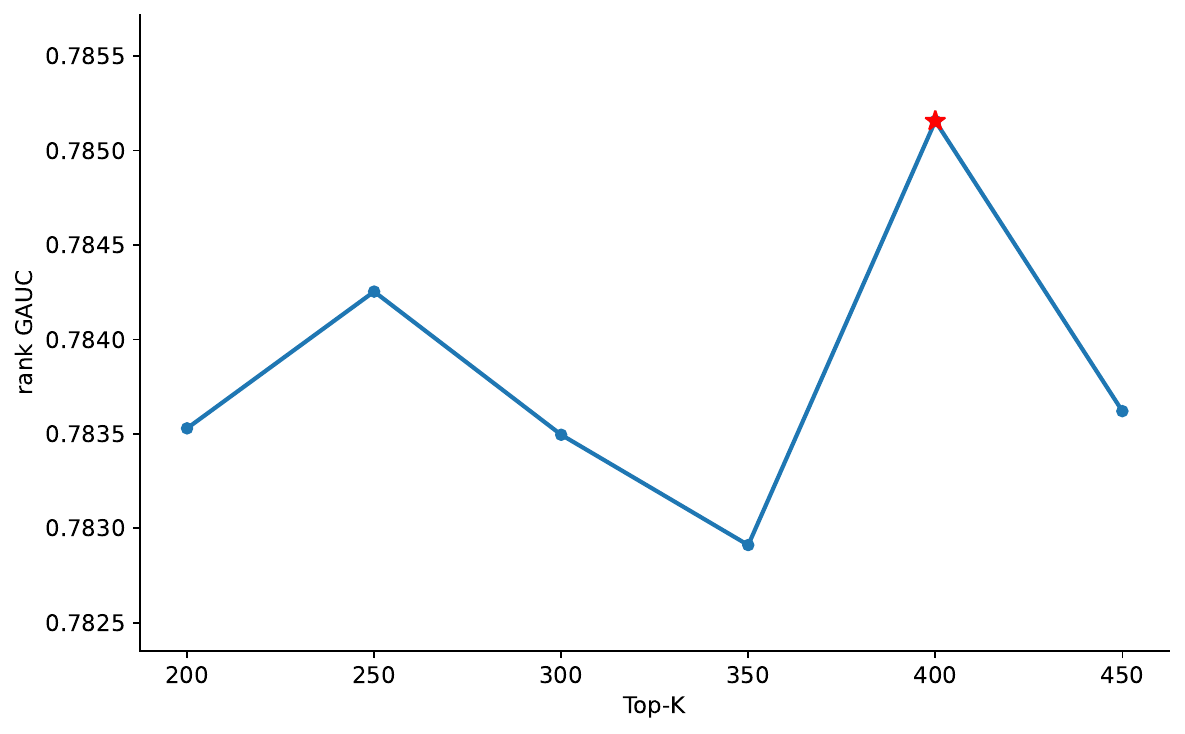}
\caption{Ranking GAUC under pair retention sizes $K$.}
\label{fig:app_cutk}
\end{figure}

\subsection{Parameter Sensitivity (RQ3)}
\label{sec:sensitivity}

We sweep CAPP's pair retention size $K$ (top pairs by confidence $c_{ij}$, Fig.~\ref{fig:app_cutk}), finding $K{=}400$ optimal. We sweep $\lambda_{\mathrm{align}}$ and the two ranking-stage weights $\lambda_{\mathrm{fb}}$ and $\lambda_{\mathrm{AGRR}}$ in Appendix Figs.~\ref{fig:app_align}--\ref{fig:app_grpo}; the pre-ranking weights $\lambda_{\mathrm{fb}}^{pre}$ and $\lambda_{\mathrm{AGRR}}^{pre}$ show similar trends and are omitted. The curves are stable around the selected values, which we set to $\lambda_{\mathrm{align}}{=}1$, $\lambda_{\mathrm{fb}}{=}1$, and $\lambda_{\mathrm{AGRR}}{=}0.3$. Excessive weights degrade performance; e.g., rank\_GAUC drops to 0.7692 as $\lambda_{\mathrm{align}}$ increases from 1 to 5, as overly strong alignment forces the upstream score toward the downstream ordering at the expense of its own objective.

\subsection{Online A/B Test Experiment (RQ4)}
\label{sec:online}
\begin{figure}[htbp]
\centering
\includegraphics[width=1.0\columnwidth]{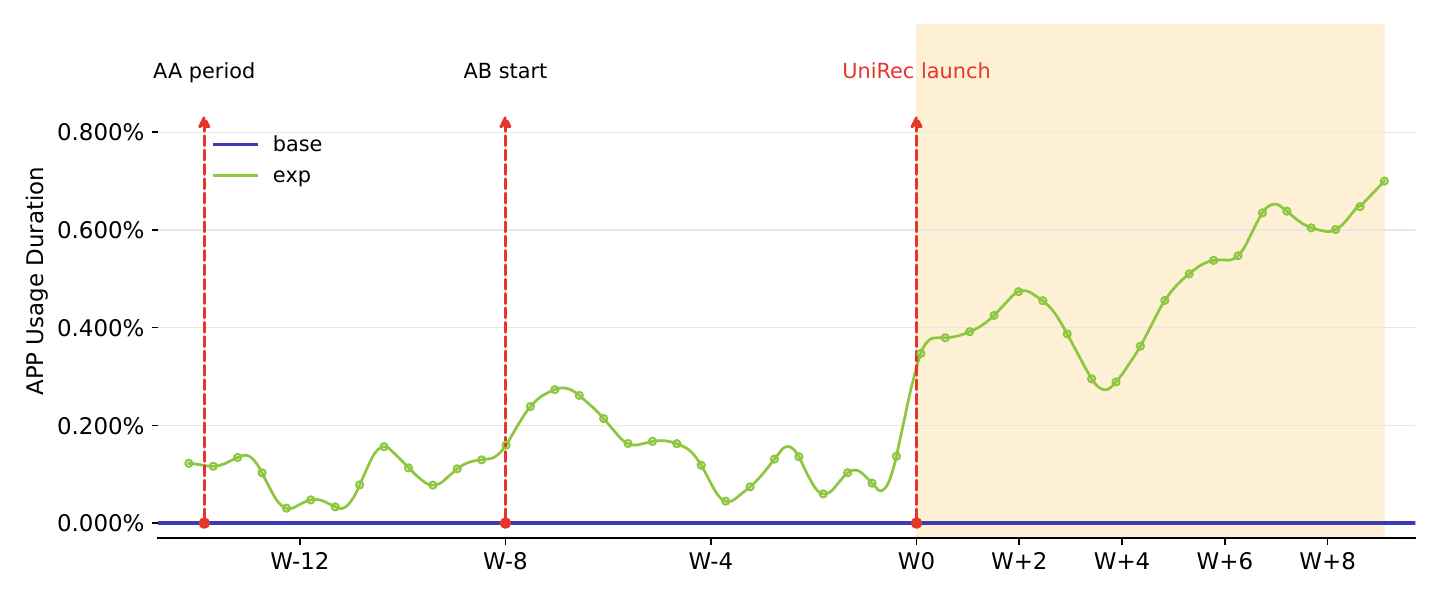}
\caption{Online A/B test: app usage time.}
\label{fig:ab_app_duration}
% \vspace{-10pt}
\end{figure}

We answer RQ4 through an online A/B test in Kuaishou's recommendation system, where the treatment group replaced the fusion modules at both stages while keeping all other components unchanged. The experiment ran on 5\% of randomly selected live traffic and has been monitored for nine weeks since launch. App usage time increased and then remained stable rather than decaying over the observation period (Fig.~\ref{fig:ab_app_duration}); we report the metrics from the final week. Compared with the production fusion baseline (Table~\ref{tab:online_ab}), UniRec improves total watch time by +0.675\%, video watch time by +0.755\%, app usage time by +0.616\%, and active users by +0.189\%, with no reported metric declining. These results indicate that the gains do not come from trading one objective against another. UniRec has now been fully deployed in Kuaishou's recommendation system.

\begin{table}[htbp]
\centering
\small
\setlength{\tabcolsep}{6pt}
\caption{Online A/B results.}
\label{tab:online_ab}
\begin{tabular}{lr@{\hspace{40pt}}lr}
\toprule
Metric & Rel. impr. & Metric & Rel. impr. \\
\midrule
LT7 & +0.142\% & Realshow & +0.427\% \\
ActiveUsers & +0.189\% & LongView & +1.028\% \\
AppUsageTime & +0.616\% & Like & +2.367\% \\
TotalWatchTime & +0.675\% & Follow & +1.573\% \\
VideoWatchTime & +0.755\% & Collect & +1.890\% \\
\bottomrule
\end{tabular}
\end{table}

\section{Conclusion}
We identify cross-stage preference inconsistency in multi-task fusion as an important challenge in cascaded recommender systems and propose UniRec, a unified framework that jointly optimizes pre-ranking and ranking fusion agents. UniRec couples the two stages through partially shared embeddings and cross-stage preference optimization while preserving stage-specific fusion functions. It further introduces AGRR to control reward-induced distribution drift and CPPA with CAPP to reduce the computational complexity of multi-stage pairwise optimization. Extensive offline and large-scale online experiments on a production short-video system demonstrate consistent gains over cross-stage coordination and single-stage fusion baselines. In a nine-week online A/B test, UniRec improves total watch time by +0.675\%, app usage time by +0.616\%, and active users by +0.189\%, which has been fully deployed in Kuaishou's recommendation system.

\clearpage
\bibliographystyle{ACM-Reference-Format}
\bibliography{reference}

@inproceedings{cao2025xmtf,
  title={xMTF: A Formula-Free Model for Reinforcement-Learning-Based Multi-Task Fusion in Recommender Systems},
  author={Cao, Yang and Zhang, Changhao and Chen, Xiaoshuang and Zhan, Kaiqiao and Wang, Ben},
  booktitle={Proceedings of the ACM on Web Conference 2025},
  pages={3840--3849},
  year={2025}
}

@article{lowe2017multi,
  title={Multi-agent actor-critic for mixed cooperative-competitive environments},
  author={Lowe, Ryan and Wu, Yi I and Tamar, Aviv and Harb, Jean and Pieter Abbeel, OpenAI and Mordatch, Igor},
  journal={Advances in neural information processing systems},
  volume={30},
  year={2017}
}

@inproceedings{lin2019pareto,
  title={A pareto-efficient algorithm for multiple objective optimization in e-commerce recommendation},
  author={Lin, Xiao and Chen, Hongjie and Pei, Changhua and Sun, Fei and Xiao, Xuanji and Sun, Hanxiao and Zhang, Yongfeng and Ou, Wenwu and Jiang, Peng},
  booktitle={Proceedings of the 13th ACM Conference on recommender systems},
  pages={20--28},
  year={2019}
}

@inproceedings{tang2020progressive,
  title={Progressive layered extraction (ple): A novel multi-task learning (mtl) model for personalized recommendations},
  author={Tang, Hongyan and Liu, Junning and Zhao, Ming and Gong, Xudong},
  booktitle={Proceedings of the 14th ACM conference on recommender systems},
  pages={269--278},
  year={2020}
}

@incollection{wang2017deep,
  title={Deep \& cross network for ad click predictions},
  author={Wang, Ruoxi and Fu, Bin and Fu, Gang and Wang, Mingliang},
  booktitle={Proceedings of the ADKDD'17},
  pages={1--7},
  year={2017}
}

@inproceedings{ma2018modeling,
  title={Modeling task relationships in multi-task learning with multi-gate mixture-of-experts},
  author={Ma, Jiaqi and Zhao, Zhe and Yi, Xinyang and Chen, Jilin and Hong, Lichan and Chi, Ed H},
  booktitle={Proceedings of the 24th ACM SIGKDD international conference on knowledge discovery \& data mining},
  pages={1930--1939},
  year={2018}
}

@article{tolstikhin2021mlp,
  title={Mlp-mixer: An all-mlp architecture for vision},
  author={Tolstikhin, Ilya O and Houlsby, Neil and Kolesnikov, Alexander and Beyer, Lucas and Zhai, Xiaohua and Unterthiner, Thomas and Yung, Jessica and Steiner, Andreas and Keysers, Daniel and Uszkoreit, Jakob and others},
  journal={Advances in neural information processing systems},
  volume={34},
  pages={24261--24272},
  year={2021}
}

@inproceedings{song2019autoint,
  title={Autoint: Automatic feature interaction learning via self-attentive neural networks},
  author={Song, Weiping and Shi, Chence and Xiao, Zhiping and Duan, Zhijian and Xu, Yewen and Zhang, Ming and Tang, Jian},
  booktitle={Proceedings of the 28th ACM international conference on information and knowledge management},
  pages={1161--1170},
  year={2019}
}

@inproceedings{cao2025pantheon,
  title={Pantheon: Personalized multi-objective ensemble sort via iterative pareto policy optimization},
  author={Cao, Jiangxia and Xu, Pengbo and Cheng, Yin and Guo, Kaiwei and Tang, Jian and Wang, Shijun and Leng, Dewei and Yang, Shuang and Liu, Zhaojie and Niu, Yanan and others},
  booktitle={Proceedings of the 34th ACM International Conference on Information and Knowledge Management},
  pages={5575--5582},
  year={2025}
}

@article{sener2018multi,
  title={Multi-task learning as multi-objective optimization},
  author={Sener, Ozan and Koltun, Vladlen},
  journal={Advances in neural information processing systems},
  volume={31},
  year={2018}
}

@article{xia2026harmonrank,
  title={HarmonRank: Ranking-aligned Multi-objective Ensemble for Live-streaming E-commerce Recommendation},
  author={Xia, Boyang and Yu, Zhou and Zhu, Zhiliang and Sun, Hanxiao and Han, Biyun and Wang, Jun and Liu, Runnan and Ou, Wenwu},
  journal={arXiv preprint arXiv:2601.02955},
  year={2026}
}

@inproceedings{zhang2022multi,
  title={Multi-task fusion via reinforcement learning for long-term user satisfaction in recommender systems},
  author={Zhang, Qihua and Liu, Junning and Dai, Yuzhuo and Qi, Yiyan and Yuan, Yifan and Zheng, Kunlun and Huang, Fan and Tan, Xianfeng},
  booktitle={Proceedings of the 28th ACM SIGKDD conference on knowledge discovery and data mining},
  pages={4510--4520},
  year={2022}
}

@inproceedings{zhang2024unex,
  title={UNEX-RL: reinforcing long-term rewards in multi-stage recommender systems with unidirectional execution},
  author={Zhang, Gengrui and Wang, Yao and Chen, Xiaoshuang and Qian, Hongyi and Zhan, Kaiqiao and Wang, Ben},
  booktitle={Proceedings of the AAAI Conference on Artificial Intelligence},
  volume={38},
  number={8},
  pages={9305--9313},
  year={2024}
}

@inproceedings{zhao2025hybrid,
  title={A hybrid cross-stage coordination pre-ranking model for online recommendation systems},
  author={Zhao, Binglei and Qi, Houying and Xu, Guang and Ma, Mian and Zhao, Xiwei and Mei, Feng and Xu, Sulong and Hu, Jinghe},
  booktitle={Companion Proceedings of the ACM on Web Conference 2025},
  pages={621--630},
  year={2025}
}

@article{oord2018representation,
  title={Representation learning with contrastive predictive coding},
  author={Oord, Aaron van den and Li, Yazhe and Vinyals, Oriol},
  journal={arXiv preprint arXiv:1807.03748},
  year={2018}
}

@inproceedings{cai2023reinforcing,
  title={Reinforcing user retention in a billion scale short video recommender system},
  author={Cai, Qingpeng and Liu, Shuchang and Wang, Xueliang and Zuo, Tianyou and Xie, Wentao and Yang, Bin and Zheng, Dong and Jiang, Peng and Gai, Kun},
  booktitle={Companion Proceedings of the ACM Web Conference 2023},
  pages={421--426},
  year={2023}
}

@inproceedings{zhao2023copr,
  title={COPR: consistency-oriented pre-ranking for online advertising},
  author={Zhao, Zhishan and Gao, Jingyue and Zhang, Yu and Han, Shuguang and Lou, Siyuan and Sheng, Xiang-Rong and Wang, Zhe and Zhu, Han and Jiang, Yuning and Xu, Jian and others},
  booktitle={Proceedings of the 32nd ACM International Conference on Information and Knowledge Management},
  pages={4974--4980},
  year={2023}
}

@article{zhang2023rethinking,
  title={Rethinking the role of pre-ranking in large-scale e-commerce searching system},
  author={Zhang, Zhixuan and Huang, Yuheng and Ou, Dan and Li, Sen and Li, Longbin and Liu, Qingwen and Zeng, Xiaoyi},
  journal={arXiv preprint arXiv:2305.13647},
  year={2023}
}

@article{rubinstein1999cross,
  title={The cross-entropy method for combinatorial and continuous optimization},
  author={Rubinstein, Reuven},
  journal={Methodology and computing in applied probability},
  volume={1},
  number={2},
  pages={127--190},
  year={1999},
  publisher={Springer}
}

@inproceedings{movckus1974bayesian,
  title={On Bayesian methods for seeking the extremum},
  author={Mo{\v{c}}kus, Jonas},
  booktitle={IFIP Technical Conference on Optimization Techniques},
  pages={400--404},
  year={1974},
  organization={Springer}
}

@article{wang2020cold,
  title={Cold: Towards the next generation of pre-ranking system},
  author={Wang, Zhe and Zhao, Liqin and Jiang, Biye and Zhou, Guorui and Zhu, Xiaoqiang and Gai, Kun},
  journal={arXiv preprint arXiv:2007.16122},
  year={2020}
}

@article{xu2025umre,
  title={UMRE: A Unified Monotonic Transformation for Ranking Ensemble in Recommender Systems},
  author={Xu, Zhengrui and Yang, Zhe and Guo, Zhengxiao and Liu, Shukai and Lin, Luocheng and Liu, Xiaoyan and Liu, Yongqi and Li, Han},
  journal={arXiv preprint arXiv:2508.07613},
  year={2025}
}

@article{he2025end,
  title={An End-to-End Multi-objective Ensemble Ranking Framework for Video Recommendation},
  author={He, Tiantian and Xie, Minzhi and Li, Runtong and Xu, Xiaoxiao and Yu, Jiaqi and Wang, Zixiu and Hu, Lantao and Li, Han and Gai, Kun},
  journal={arXiv preprint arXiv:2508.05093},
  year={2025}
}

@article{shao2024deepseekmath,
  title={Deepseekmath: Pushing the limits of mathematical reasoning in open language models},
  author={Shao, Zhihong and Wang, Peiyi and Zhu, Qihao and Xu, Runxin and Song, Junxiao and Bi, Xiao and Zhang, Haowei and Zhang, Mingchuan and Li, YK and Wu, Yang and others},
  journal={arXiv preprint arXiv:2402.03300},
  year={2024}
}

@article{liu2024recflow,
  title={RecFlow: An Industrial Full Flow Recommendation Dataset},
  author={Liu, Qi and Zhao, Kai and Xie, Ruiwen and others},
  journal={arXiv preprint arXiv:2410.20868},
  year={2024}
}

@inproceedings{covington2016deep,
  title={Deep neural networks for YouTube recommendations},
  author={Covington, Paul and Adams, Jay and Sargin, Emre},
  booktitle={Proceedings of the 10th ACM Conference on Recommender Systems},
  pages={191--198},
  year={2016}
}

\clearpage
\appendix

\section{Hyper-parameter Sensitivity}
\label{app:sensitivity}
In each sweep only the target coefficient varies; all remaining coefficients, the backbone, the data slice and the random seed are held fixed. The red star marks the setting adopted throughout the paper.

\begin{figure}[H]
\centering
\includegraphics[width=0.8\columnwidth]{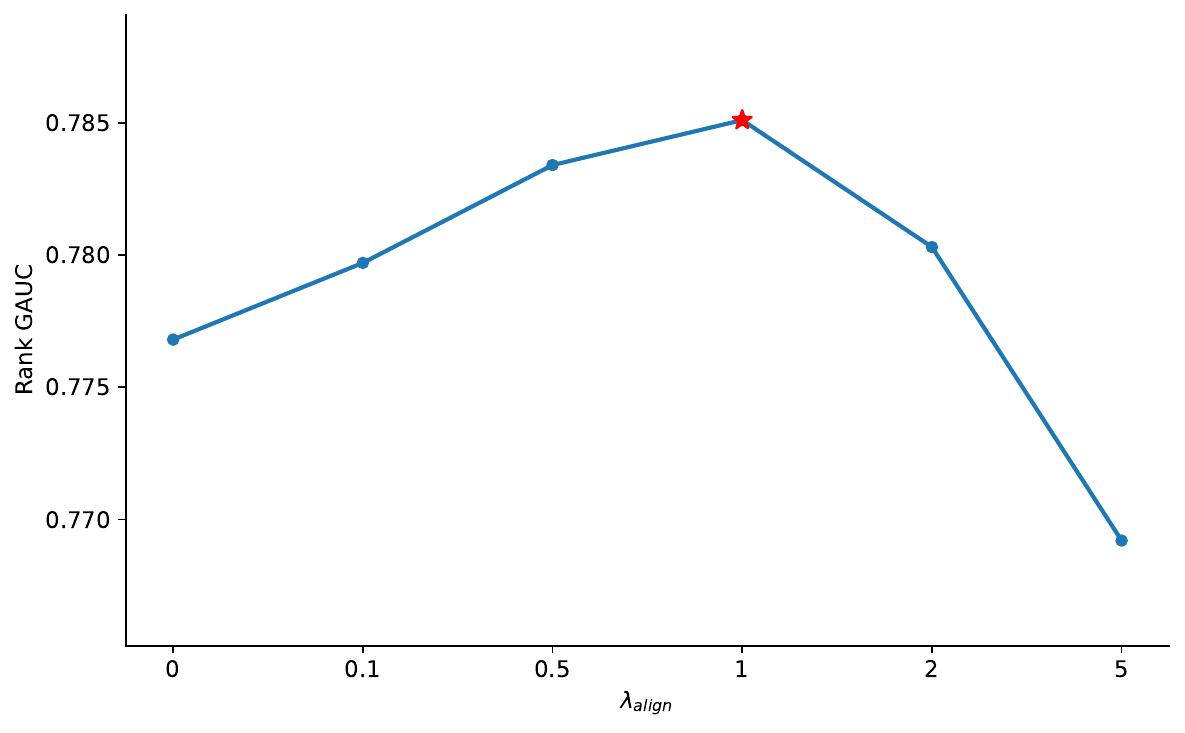}
\caption{Ranking GAUC under alignment loss weights $\lambda_{\text{align}}$.}
\label{fig:app_align}
\end{figure}

\begin{figure}[H]
\centering
\includegraphics[width=0.8\columnwidth]{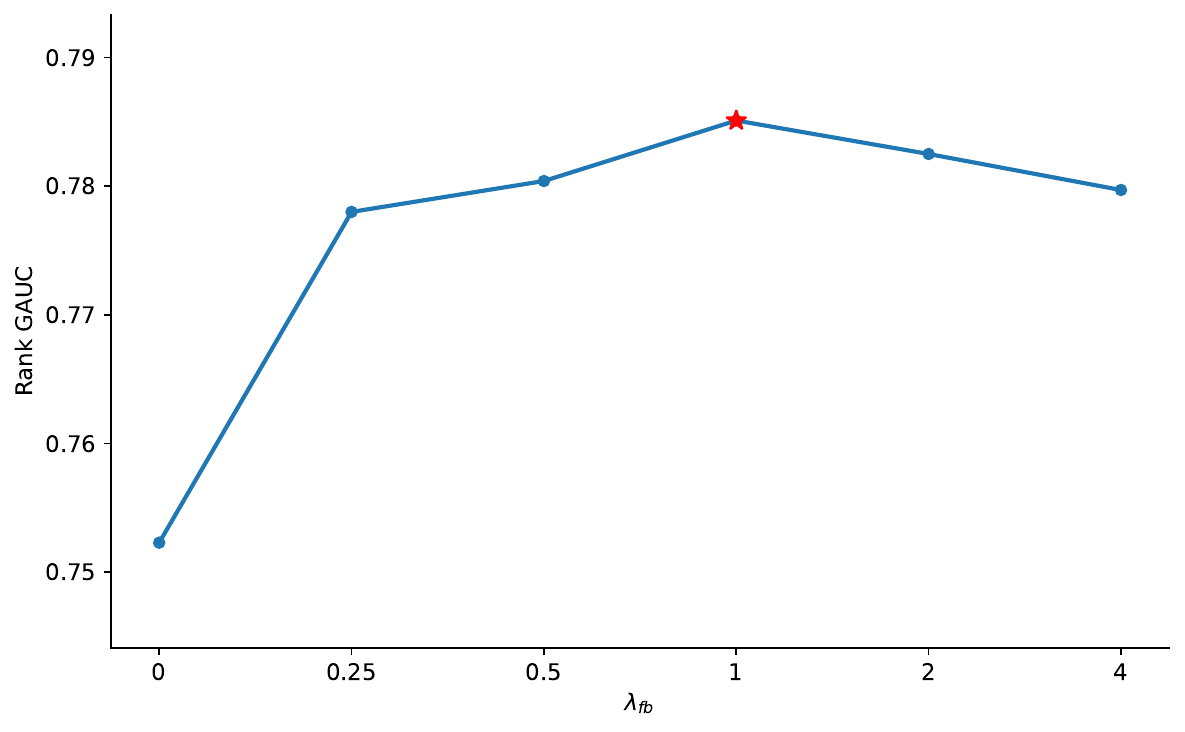}
\caption{Ranking GAUC under feedback loss weights $\lambda_{\text{fb}}$.}
\label{fig:app_fb}
\end{figure}

\begin{figure}[H]
\centering
\includegraphics[width=0.8\columnwidth]{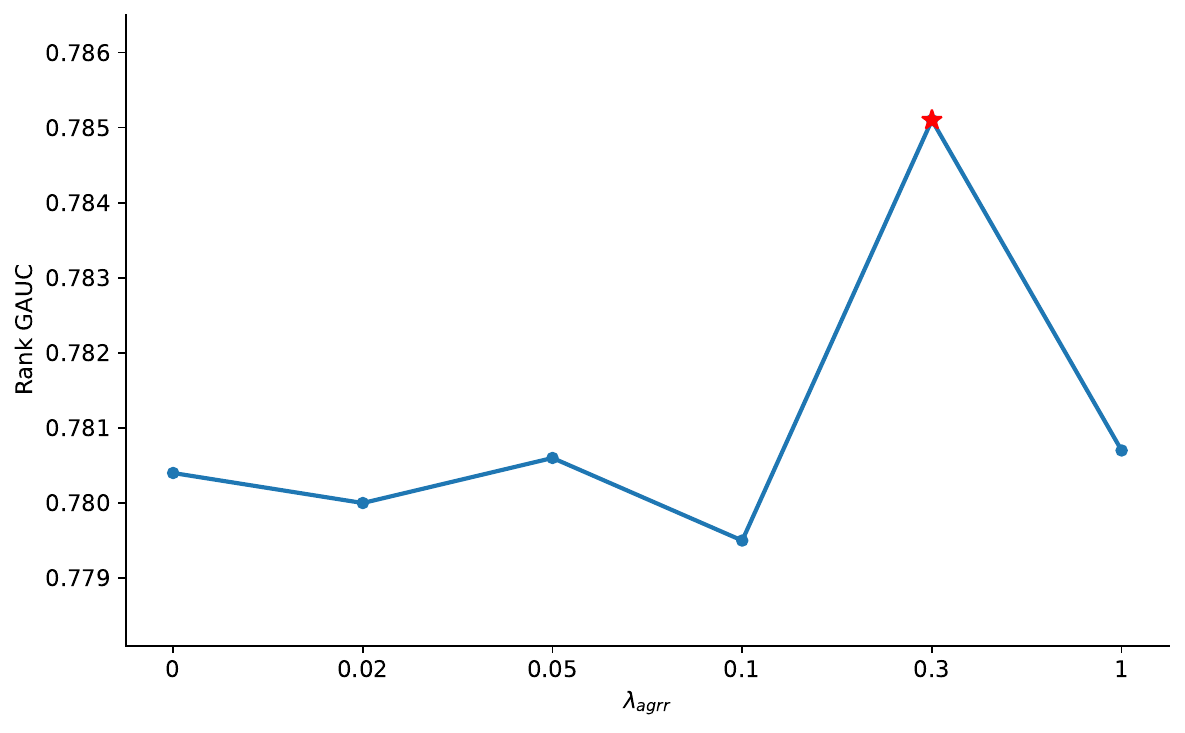}
\caption{Ranking GAUC under AGRR strengths $\lambda_{\text{AGRR}}$.}
\label{fig:app_grpo}
\end{figure}

\section{Computational Complexity Optimization}

\begin{figure}[htbp]
\centering
\includegraphics[width=0.8\columnwidth]{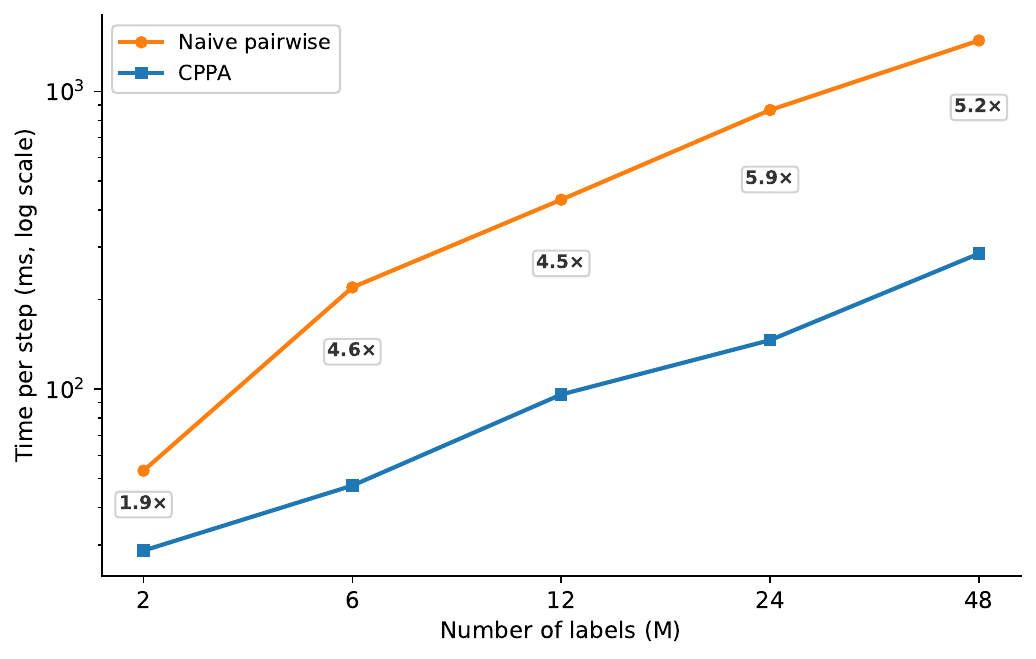}
\caption{Pairwise loss cost scaling with label count $M$.}
\label{app_cppa_scale}
\end{figure}

\begin{figure}[htbp]
\centering
\includegraphics[width=0.8\columnwidth]{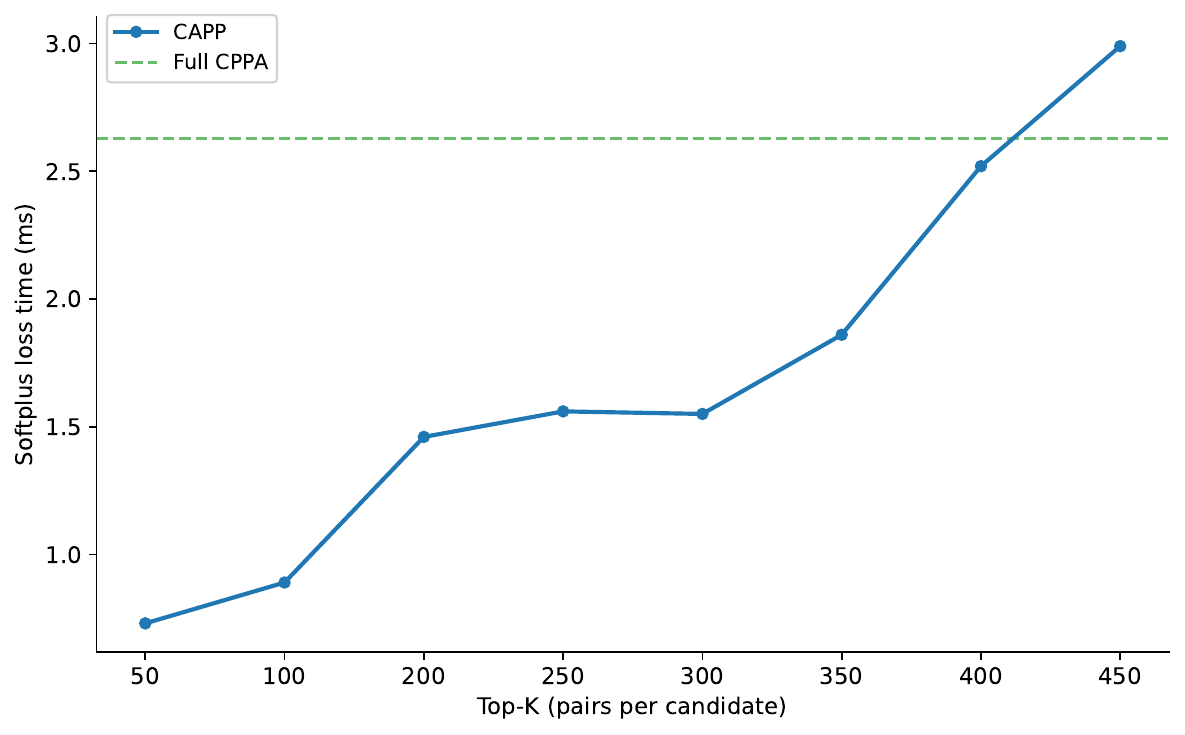}
\caption{Pairwise loss cost under pair retention size $K$.}
\label{app_diff_k}
\end{figure}

We compare UniRec (CPPA) against UniRec-NoCPPA, which replaces CPPA with naive pairwise loss. Table~\ref{tab:efficiency} reports four profiling granularities :  loss\_fn (rank/pre) measures the isolated pairwise loss computation on pre-computed embeddings for each stage; compute\_loss adds the full encoder forward pass (both stages) plus all auxiliary loss terms (posterior feedback,  alignment, AGRR); train\_step covers the complete optimization step (compute\_loss $+$ {zero\_grad} $+$ backward $+$ optimizer). Throughput reports the number of training samples processed per second. Figure~\ref{app_cppa_scale} shows how training time scales with the number of labels.
Figure~\ref{app_diff_k} reports the isolated softplus loss time under the pair
pruning: at $K{=}50$, the $\mathcal{O}(KN)$
evaluation is $3.6\times$ faster than full pairwise ($\mathcal{O}(N^2)$); as
$K \to N$ the cost converges to the full UniRec baseline.

\section{AGRR Online Duration Distribution}
\begin{figure}[H]
\centering
\includegraphics[width=0.95\columnwidth]{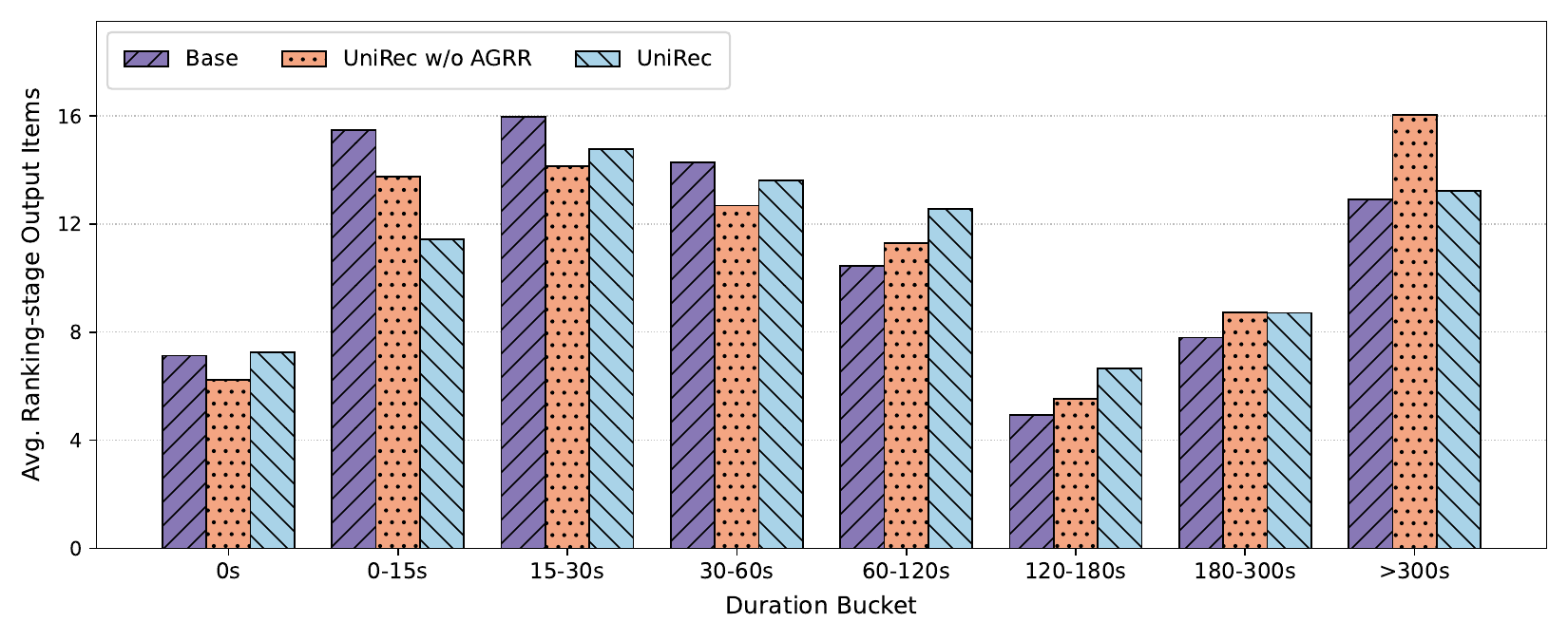}
\caption{Online duration distribution comparison between UniRec and UniRec w/o AGRR.}
\label{agrr_duration}
\end{figure}

We compare the online duration distribution of ranking-stage output items across the production base, UniRec, and UniRec w/o AGRR. Figure~\ref{agrr_duration} reports the average item count per duration
bucket. 

\end{document}